\documentclass[preprint,preprintnumbers,12pt]{elsarticle}

\usepackage{amssymb}
\usepackage{amsmath}
\usepackage{graphicx}
\usepackage{dcolumn}
\usepackage{bm}
\usepackage{braket}
\usepackage{hyperref}

\newcommand{\be}{\begin{equation}}
\newcommand{\ee}{\end{equation}}
\newcommand{\fn}{\footnote}
\newcommand{\LQ}{\Lambda_{\rm QCD}}

\newcommand{\lt}{\left}
\newcommand{\rt}{\right}

\newcommand{\mf}{\mu_f}

\newcommand{\US}{\rm US}

\newcommand{\DV}{\Delta V}
\newcommand{\LMS}{\Lambda_{\overline{\rm MS}}}

\journal{Physics Letter B}

\begin{document}


\begin{frontmatter}



\title{Determination of $\alpha_s$ from static QCD potential\\
 with renormalon subtraction}


\author{H.~Takaura$^a$, T.~Kaneko$^b$, Y.~Kiyo$^c$ and Y.~Sumino$^d$}

\address{
\mbox{$^a$Department of Physics, Kyushu University, Fukuoka, 819--0395 Japan}
\\
$^b$Theory Center, KEK, Tsukuba, Ibaraki, 305--0801 Japan
\\
\mbox{$^c$Department of Physics, Juntendo University, Inzai, 270--1695 Japan}
\\
\mbox{$^d$Department of Physics, Tohoku University,
Sendai, 
980--8578 Japan}
}

\begin{abstract}
We determine the strong coupling constant $\alpha_s(M_Z)$ from the static QCD potential
by matching a lattice result and a theoretical calculation.
We use a new theoretical framework based on operator product expansion (OPE),
where renormalons are subtracted from the leading Wilson coefficient.
We find that our OPE prediction can explain the lattice data at $\LQ r \lesssim 0.8$.
This allows us to use a larger window in matching, which leads to a more reliable determination.
We obtain $\alpha_s(M_Z)=0.1179^{+0.0015}_{-0.0014}$.
\end{abstract}






\end{frontmatter}

Today, facing frontier experiments of particle physics
such as those at LHC and
super $B$ Factory, there exist increasing demands for more
accurate theoretical predictions
based on QCD on various phenomena of the strong interaction.
Precise determination of the strong coupling
constant $\alpha_s$ sets a benchmark for such predictions.
For instance, a precise value of $\alpha_s$ 
will play crucial roles in measurements of Higgs boson properties, 
in searches for new physics, in high-precision flavor physics, etc.

The current value of $\alpha_s$,
given as the world-combined result by the Particle Data Group (PDG), 
reads
$
\alpha_s(M_Z)=0.1181 \pm 0.0011 
$
\cite{37ce3e5843594be4beddf3c7540d08bc}.
Dominant contributions to this value come from
determinations by lattice QCD,
which have smaller errors than other determinations using 
more direct experimental inputs.
Nevertheless, 
most lattice QCD determinations have 
the ``window problem" in an explicit or implicit way,
as pointed out in the Flavor Lattice Averaging Group (FLAG) report \cite{Aoki:2016frl}: 
It is difficult to find a region where 
both lattice QCD and perturbative QCD predictions are accurate.
At short distances ($Q \gg \LQ$), where perturbation 
theory is accurate, lattice data  
are distorted by ultraviolet (UV) cutoff effects due to the finite
lattice spacing $a$,
whereas at larger distances ($Q \sim \LQ$), where finite $a$ effects are 
suppressed,
perturbation theory is not reliable.  

The method of the finite volume scheme combined with
step-scaling was proposed to solve this problem \cite{Luscher:1991wu,Bruno:2017gxd}.
This method enlarges reliable energy region of lattice simulation.
As a result, matching with perturbative prediction can be taken 
in a wide range at high energy, 10--100~{\rm GeV}.

In this Letter, we propose an alternative approach to the window problem:
We enlarge the validity range of theoretical prediction 
to the region where lattice calculations are accurate, $Q \ll a^{-1}$.
To this end we use operator product expansion (OPE) 
with subtraction of renormalons.
Accuracy of a perturbative prediction has a limitation due to renormalons 
(which specify certain divergent behaviors of perturbative series),
and an $\mathcal{O}(\LQ^n/Q^n)$-error is inevitable, for a dimensionless observable
with typical scale $Q$.
In OPE, the $\mathcal{O}(\LQ^n/Q^n)$-term is described 
by a nonperturbative matrix element (ME).
Hence, we can enlarge validity range
of theoretical prediction to lower energy
by subtracting renormalons appropriately from perturbative
prediction in the framework of
OPE.

Although OPE is a good and well-known framework, 
there is a difficulty in practical calculations.
There has not been an established way to factorize the two components of OPE, Wilson coefficients and nonperturbative MEs,
which are conceptually UV and  infrared (IR) quantities, respectively.
Although one may find in the literature that Wilson coefficients are calculated in usual perturbation theory,
this procedure is not desirable since loop integrals in dimensional regularization contain both UV and IR modes. 
In particular, IR contributions cause renormalon uncertainties in a Wilson coefficient,
which makes it practically  impossible to distinguish a nonperturbative ME 
from the renormalon uncertainties.

In Refs.~\cite{Sumino:2005cq, Mishima:2016vna},  a formulation to separate UV and IR contributions
in OPE has been proposed.  
A Wilson coefficient is constructed as a UV quantity,
free from renormalon uncertainties
(i.e., renormalons are subtracted from the Wilson coefficient).
The formulation concurrently defines a nonperturbative ME as an IR quantity.
This prevents mixing of the nonperturbative ME with
renormalon uncertainties in
the Wilson coefficient,
thus enabling us to perform OPE in an ideal way.
In particular, for the static QCD potential, 
one can calculate a Wilson coefficient systematically from the fixed-order perturbative result.
It is identified with
the leading term of OPE 
in the solid framework of potential nonrelativistic QCD (pNRQCD)
effective field theory \cite{Brambilla:1999xf}.

We determine $\alpha_s$ from the QCD potential by matching 
a lattice result and OPE.
We can take
the matching range down to relatively low energy scale
$\LQ r \lesssim 0.6$--$0.8$ by subtraction of renormalons.
This is in contrast to previous $\alpha_s$ determinations using the 
QCD potential \cite{Bazavov:2014soa,Karbstein:2018mzo}, in which
lattice results are matched with perturbative 
results in the region $\LQ r \lesssim 0.2$--$0.3$.

We use the lattice result at cutoffs up to 4.5 GeV
obtained by the JLQCD collaboration.
Our theoretical calculation is based on OPE with renormalon subtraction 
and the 
next-to-next-to-next-to-leading order (${\rm N^3}$LO) result of perturbation theory.
The unique feature of our method is to perform OPE avoiding the mixing of
a Wilson coefficient and a nonperturbative ME.
This clarifies their respective roles,
and an estimate of theoretical error can be given clearly.
In contrast, in many studies considering OPE, 
an estimate of perturbative error cannot be distinguished from an estimate of
nonperturbative effects, since they are mixed in a naive calculation method.
We will show that our OPE prediction can explain lattice data at 
$r^{-1} \gtrsim 0.5~{\rm GeV}$ (or $r \lesssim 0.4~{\rm fm}$),
where usual perturbation theory cannot work sufficiently.

Our theoretical prediction for the QCD potential
is based on multipole expansion within pNRQCD,
which is an OPE in $\vec{r}$.
The QCD potential is expanded as \cite{
Brambilla:1999xf}
\be
V_{\rm QCD}(r)=V_S(r)+\delta E_{\US}(r)+\dots \, , \label{multi}
\ee 
where the explicit $r$ dependence of each term is $V_S(r) \sim \frac{1}{r}$ and 
$\delta E_{\US}(r) \sim r^2$, and the dots denote the higher-order terms in $r$.
(We suppress the $r$-independent part.)
In the following we consider the first two terms of OPE,
shown explicitly in Eq.~(\ref{multi}), unless stated otherwise.  
While the singlet potential
$V_S$ is a UV quantity, $\delta E_{\US}$ and higher correction terms 
are dominantly 
IR quantities determined by nonperturbative dynamics.
In usual perturbative evaluation of $V_S$, renormalon uncertainties appear,
whose leading $r$-dependent uncertainty is $\mathcal{O}(\LQ^3 r^2)$.
Ref.~\cite{Brambilla:1999xf} has pointed out that this renormalon uncertainty
is canceled against that of $\delta E_{\US}$.
This observation suggests that one should subtract renormalons from $V_S$ 
to define it as an unambiguous object and also to make $\delta E_{\US}$ free from renormalons.

We subtract renormalon uncertainties of $V_S$ as follows \cite{Sumino:2005cq, Mishima:2016vna}. 
The QCD potential is formally given by
\be
V_{\rm QCD}(r)=-4  \pi C_F \! \int \! 
\frac{d^3 \vec{q}}{(2 \pi)^3} \, e^{i \vec{q} \cdot \vec{r} } \, 
\frac{\alpha_V(q)}{q^2} \, .
~~~(q=|\vec{q}|)
\ee
In this expression $q$
varies from 0 to $\infty$.
Since the singlet potential corresponds to UV part of $V_{\rm QCD}$, 
we define 
\be
V_S(r;\mu_f)=-4  \pi C_F \int _{q>\mf} \frac{d^3 \vec{q}}{(2 \pi)^3} \, 
e^{i \vec{q} \cdot \vec{r} } \, \frac{\alpha_V(q)}{q^2} \label{VScutoff} \, ,
\ee
with a factorization scale $\mu_f$.
$V_S$ does not have renormalon uncertainties since IR contributions
 are removed.\footnote{
 More accurately, dominant renormalons which arise from the $\vec{q}$-integral
 are removed.
 Renormalons contained in $\alpha_V(q)$ are subdominant and have not been
 well studied, which we neglect in this analysis.
 }
In $V_S$, $\mf$-dependent part is sensitive to 
IR dynamics, and when combined with $\delta E_{\US}$,
it becomes independent of $\mu_f$ 
[up to $\mathcal{O}(r^2)$].
In contrast, $\mf$-independent part of $V_S$
corresponds to a pure UV contribution, which is 
accurately predictable within perturbation theory.

\begin{figure}[t]
\begin{center}
\includegraphics[width=7.9cm]{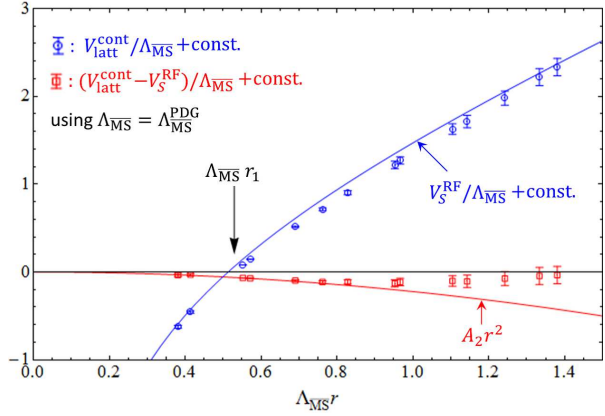}
\end{center}
\vspace*{-5mm}
\caption{
Comparison of the lattice result (cont.\ limit: blue circles) 
and leading OPE prediction ($V_S^{\rm RF}/\LMS\,$: blue line)
using $\LMS^{\rm PDG}$ and adjusting $r$-independent part.
The difference (red boxes) 
is fitted by ${\rm const.}\times r^2$ (red line) at small $r$.
}
\label{Fig:consistency}
\vspace*{-4mm}
\end{figure}

We construct  a $\mf$-independent part of $V_S(r;\mu_f)$, denoted as $V_S^{\rm RF}(r)$,
in the following manner.
We utilize the perturbative result for $\alpha_V(q)$ 
known up to $\mathcal{O}(\alpha_s^4)$ (N$^3$LO)
\cite{Smirnov:2008pn,Anzai:2009tm, Smirnov:2009fh}.
We improve the fixed-order result 
by renormalization group (RG) 
using the 4-loop $\beta$-function.
Up to here, the integrand of Eq.~(\ref{VScutoff}) is determined.
Then, by deforming the integral path in the complex-$q$ plane, 
we can extract a $\mu_f$-independent singlet potential $V_S^{\rm RF}(r)$ 
with N$^3$LL (leading log) accuracy; see Ref.~\cite{Sumino:2005cq} for details.  
$V_S^{\rm RF}$ does not have renormalon uncertainties or factorization scale dependence.
The factorization scale dependent part of $V_S$ is absorbed into
the nonperturbative ME.
By this, $\mu_f$ dependence of the nonperturbative ME vanishes 
as well \cite{Takaura:2017lwd}.
In this way, one can resolve the mixing of the Wilson coefficient $V_S$ with the nonperturbative term $\delta E_{\US}$,
and obtains
\be
V_{\rm QCD}(r)=V_S^{\rm RF}(r)+\delta E_{\rm US}^{\rm RF}(r)+\dots \, ,
\ee
where each term is free of renormalons and $\mf$.
In our analysis, we regard $\delta E_{\rm US}^{\rm RF}$ as the non-local gluon condensate\fn{
Proper treatment of $\delta E_{\rm US}$ depends on distance region:
it is a perturbative contribution when the ultrasoft scale $\Delta V(r)=C_A \alpha_s /(2 r)$ (with $C_A=3$) satisfies $\Delta V \gg \LQ$,
while it is a nonperturbative condensate when  $\Delta V \lesssim \LQ$.
Although $\DV \gg \LQ$ is satisfied at very short distances, 
$\DV$ and $\LQ$ have similar sizes at $\LQ r \gtrsim 0.2$.
Since our fitting range extends to relatively long distances $\LQ r< 0.6$--0.8,
we regard $\delta E_{\rm US}$ as a nonperturbative contribution. 
(Validity of this treatment is shown in Fig.~\ref{Fig:consistency}.)}
of order $\LQ^3 r^2$.

The OPE prediction is compared to the potential $V_{\rm latt}$ 
calculated nonperturbatively in 3-flavor lattice QCD
in the isospin limit~\cite{Kaneko:2013jla}.
%
%
We employ the Symanzik gauge~\cite{Weisz:1982zw}
and M\"obius domain-wall quark actions~\cite{Kaneko:2013jla,Brower:2012vk},
with which the leading discretization effect is $O(a^2)$.
The lattice simulations are carried out at three lattice cutoffs,
determined as $a^{-1}\!=\!2.453(4)$, 3.610(9) and 4.496(9)~GeV
from the Wilson-flow scale \cite{Luscher:2010iy}.
The 
lattice sizes at these cutoffs are
$32^3\!\times\!64$, $48^3\!\times\!96$ and $64^3\!\times\!128$, respectively,
with the physical size roughly kept fixed.
At each $a^{-1}$,
we take a single combination of the light and strange quark masses
$(m_{ud}^{\rm latt},m_s^{\rm latt})$, which roughly correspond to
$(M_\pi,M_K)\!\sim\!(300\,\mbox{MeV},520\,\mbox{MeV})$.
%
%
We extract $V_{\rm latt}$ from Wilson loops
with the spatial Wilson lines parallel to
the spatial directions $(1,0,0)$ and $(1,1,0)$,
denoted as directions 1 and 2, respectively.

We determine $\alpha_s$ with the following strategy, 
by performing two analyses with different methods. 
The first analysis [Analysis (I)] consists of two steps: extracting a continuum limit of the lattice result
and determination of $\alpha_s$ by comparing the OPE prediction with the continuum limit. 
We proceed while checking (i) if the lattice data can be smoothly 
extrapolated to the continuum limit,
and (ii) if $V_S^{\rm RF}(r)$  can explain the lattice result 
up to nonperturbative effects
of $\mathcal{O}(r^2)$.
After confirming these features, 
we perform a global fit to determine $\alpha_s$ in the second analysis [Analysis (II)],
without separating continuum extrapolation of the lattice data and extraction of $\alpha_s$.   
Analysis (II) is a first-principle analysis, which avoids introducing
a model interpolating function, 
required in the first analysis for continuum extrapolation.
Our final result will be adopted from Analysis (II), 
whose errors are well controlled and are smaller than that of Analysis (I).
Analysis (I) makes up for a shortcoming of Analysis (II) that the
output follows from the inputs without revealing 
detailed profiles at intermediate steps.
Throughout our analyses,
correlations among the lattice data are taken into account
by the covariant matrices and the jackknife method.
(See Ref.~\cite{prep} for the details of the analyses.)

\vspace*{1mm}
\noindent{\bf Analysis (I)}
We first extract the continuum limit of the lattice data, specifically 
$X_{\rm latt}(r) \equiv  r_1 [V_{\rm latt}(r)-V_{\rm latt}(r_1)]$, where $r_1$ is the scale defined by $r_1^2 \frac{d V}{d r}(r_1)=1$.
To construct a sequence of $X_{\rm latt}(r;a)$ at the same $r$ but different $a$'s,
we first interpolate the lattice data at each $a$ using the fitting form
\be
V_{{\rm latt},d,i}^{\rm Inter.}({r})=\frac{\alpha_{d,i}}{r}+c_{0,d,i}+\sigma_{d,i}\, r+\frac{c_{1,d,i}}{r^3}+c_{2,d,i}\, r^2 \label{fit} \, ,
\ee
where $d=1,2$ and $i=1,2,3$ specify
the direction and lattice spacing, respectively.
The first three terms are the Cornell potential.
The other terms are included to take into account lattice artifacts.
The $1/r^3$ term accounts for the $\mathcal{O}(a^2)$ discretization 
effect
whose mass dimension is one.
The last term similarly accounts for the finite volume effect.
Due to the lack of rotational symmetry on the lattice,
the coefficients in Eq.~(\ref{fit}) can depend on
$d$.
Hence, we interpolate the data separately for each $(d,i)$.
From the fit~(\ref{fit}),
we calculate $X_{\rm latt}(r;a_i)$ at each $a_i$
and at reference values of $r$ (in physical units)
where the coarsest lattice has the original data.

\begin{figure}[t]
\begin{center}
\includegraphics[width=8.5cm]{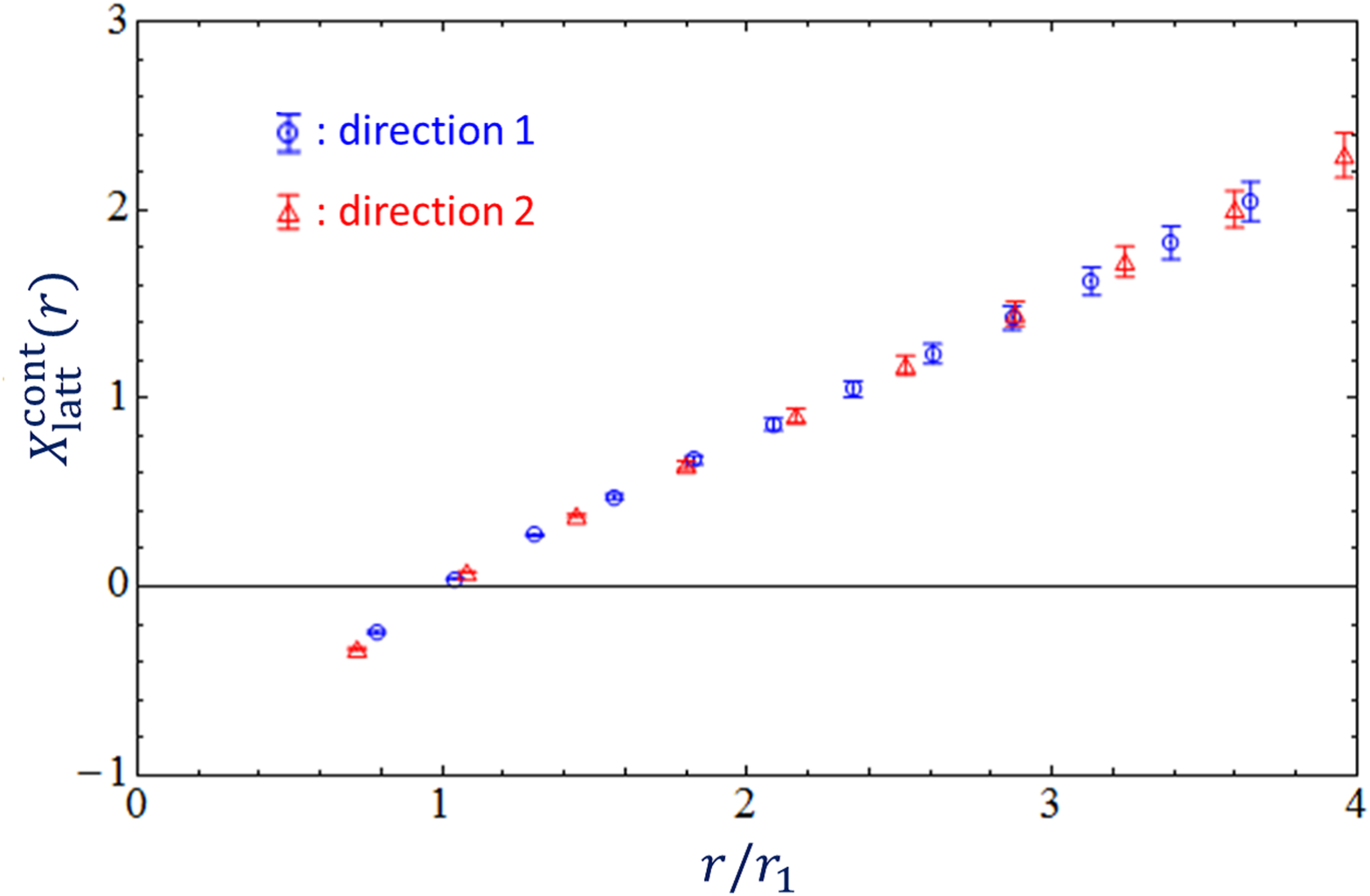}
\end{center}
\vspace*{-5mm}
\caption{Lattice result for the QCD potential after taking the continuum limit
in Analysis (I).}
\label{Fig:lattres}
\vspace*{0mm}
\end{figure}

We then extrapolate $X_{\rm latt}(r;a)$ to
$a \to 0$ by linear fits in $a^2$.
The continuum result is shown in Fig.~\ref{Fig:lattres},
where only the points which satisfy $\chi^2/{\rm d.o.f.} <2$ in
extrapolation are adopted.\footnote{
We select the lattice data at $2 a<r<L/2$ to suppress finite $a$ and $L$ effects. 
Owing to this, almost all the 
points are smoothly extrapolated to $a \to 0$ with $\chi^2/{\rm d.o.f.}<2$.
In contrast, if we include the data at $r=a$, 
$X_{\rm latt}$ does not obey a linear behavior in $a^2$,
since the interpolating function is seriously distorted by the data at $r \sim a$.
}


Before determining $\alpha_s$ from the obtained lattice result, 
as a consistency check, we confirm that
$V_S^{\rm RF}(r)$ can explain the lattice result up to 
nonperturbative corrections of  $\mathcal{O}(r^2)$.
Since the lattice result and $V_S^{\rm RF}$
are obtained in different units ($r_1$ and $\LMS^{n_f=3}$, respectively),
we need a conversion parameter $x=\LMS^{n_f=3} r_1$ to compare them.
We assume $\LMS^{n_f=3}=\LMS^{\rm PDG}=336~{\rm MeV}$ \cite{37ce3e5843594be4beddf3c7540d08bc} 
and use the central value of $r_1=0.311(2)~{\rm fm}$ \cite{Bazavov:2010hj}  
to convert the lattice result to that in $\LMS$ units. 
We see in Fig.~\ref{Fig:consistency} that the difference between the lattice data and $V_S^{\rm RF}$
can be fitted well by a constant plus an $r^2$-term
at $\LMS r \lesssim 0.8$. 
This is a first numerical observation which suggests correctness of OPE of pNRQCD in this distance range.


We also examine consistency of the lattice data at $a\!=\!0$
with other predictions:
$V_S$ in Ref.~\cite{Bazavov:2014soa} with N$^3$LL accuracy,
and the fixed-order prediction at N$^3$LO.
These contain the $\mathcal{O}(\LQ^3 r^2)$ renormalon,
and the consistency is confirmed 
in a limited range $\LMS r \lesssim 0.55$ or
strongly depends on the choice of the renormalization scale $\mu$.
In contrast,
in our formulation without the $\mathcal{O}(\LQ^3 r^2)$ renormalon,
the validity range is enlarged to $\LMS r \lesssim 0.8$
and stable against scheme choice for RG improvement, which shows an advantage of our framework.
See \cite{prep} for details.

\begin{table}[t]
\scalebox{0.67}{
\begin{tabular}{c||cccccccccc}
\hline
                                & ~~finite $a$~~            & interpol.\ fn.     & ~subt.\ point     & ~~mass~~           &  ~~~~h.o.~~ ~                  &  ~range~          &   ~~~US~~~        & fact.\ scheme  & ~$r_1$ & latt. spacing \\ \hline
Analysis (I)            &    $\pm 4$           &   $\pm 4$             &  $\pm 8$                 &  -                 &  $^{+14}_{-12}$  &  $^{+5}_{-8}$  &  $\pm1$         &  -   & $\pm 1$                            & - \\ \hline
Analysis (II)        & ~$\pm 2$              &   -                          &    -                           & $\pm 0$     & $^{+12}_{-10}$      &  $\pm 4$  &   $\pm2$         &  $\pm 3$ & ~-                                 & $\pm 4$
\\ \hline
\end{tabular}
}
\caption{Systematic errors in $\alpha_s(M_Z)$
(in units of $10^{-4}$) estimated from
the variations of $\alpha_s(M_Z)$. 
}
\label{tab:systematics}
\vspace*{-3mm}
\end{table}

Our $\alpha_s$ determination reduces to the problem 
to find an appropriate $x=\LMS^{n_f=3} r_1$ where the lattice result
agrees with the OPE prediction.
We use OPE including up to the $r^2$-term as the theoretical prediction,
\be
V_{\rm QCD}(r)=V_S^{\rm RF}(r)+A_0+A_2 r^2 \, , \label{OPE}
\ee
where $A_0$ is an $r$-independent constant
and $A_2$ specifies the size of the leading nonperturbative effect
(they are treated as fitting parameters).
We obtain $x=0.496 \pm 0.024 ({\rm stat})$ from the data at $\LMS^{\rm PDG} r<0.8$, 
adopting the range in which OPE is reliable. 
The obtained 3-flavor $\Lambda_{\overline {\rm MS}}$ gives the 5-flavor coupling 
$\alpha_s(M_Z)=0.1166^{+0.0010}_{-0.0011}({\rm stat})$
through 4-loop RG evolution 
with the charm and bottom quark threshold corrections \cite{Chetyrkin:1997sg}.
The size of the nonperturbative effect is estimated as $A_2/\Lambda_{\overline {\rm MS}}^3=0.04 \pm 0.22({\rm stat})$.

To evaluate systematic errors, we perform the following re-analyses.
(I-a) Finite $a$ effect: An analysis including shorter distance points $r > a$ is performed.
(I-b) Interpolating function: 
The fitting function (\ref{fit}) 
does not contain $\log r$ corrections (dictated by RG) in the Coulomb part 
at small $r$. 
We use another interpolating function consistent with the 1-loop RG
at small $r$.
(I-c) Subtraction point: We extract $r_1[V_{\rm latt}(r)-V_{\rm latt}(0.8 r_1)]$, where we change the subtraction point of the potential.
(I-d) Higher-order corrections to $V_S^{\rm RF}$: 
We replace $V_S^{\rm RF}$ at ${\rm N^3LL}$ by $V_S^{\rm RF} \pm \delta V_S^{\rm RF}$, where $\delta V_S^{\rm RF}$ is the difference between the ${\rm N^3LL}$
and ${\rm N^2LL}$ results.
(I-e) Matching range: To examine stability of OPE truncated at $\mathcal{O}(r^2)$,
the continuum result satisfying $\LMS^{\rm PDG} r<0.9$ or $0.7$ is used
instead of $0.8$.
(I-f) Ultrasoft (US)  contribution: $\alpha_V(q)$ at 3-loop contains an IR divergence,
which is canceled by an extra
contribution from the US scale
\cite{Appelquist:1977tw,Brambilla:1999xf}.
In the main analysis we use the LO perturbative result for
the US contribution, whereas in the error analysis we
regard the US contribution as dominantly nonperturbative and
introduce a cutoff at 
$\mu_{\US}=3\LMS$ or $4\LMS$. 
(I-g) Error of $r_1$: The scale $r_1=0.311(2)~{\rm fm}$ is varied within its error.
We summarize the systematic errors in $\alpha_s$ determination in table~\ref{tab:systematics}.

As a result of the first analysis, we obtain
\be
\alpha_s(M_Z)=
0.1166^{+0.0010}_{-0.0011}({\rm stat})^{+0.0018}_{-0.0017}({\rm sys}) \, .
\ee

\noindent{\bf Analysis (II)} 
Extrapolation to continuum limit and matching with the OPE prediction
are performed by a global fit in one step.
It is based on an idea
that the OPE prediction should coincide with the lattice result at small $r$
besides discretization effects.
The lattice data,
after correcting for discretization effects, are given by
\be
V_{{\rm latt},d,i}(r)
-\kappa_{d,i} \lt(\frac{1}{r}- \lt[\frac{1}{r} \rt]_{d,i}  \rt)
+f_d \frac{a_i^2}{r^3} - A_{0,d,i}\, .
\ee
The second term is included to remove finite-$a$ effects 
at tree level [$\kappa={\cal O}(\alpha_s)$],
where $\lt[\frac{1}{r} \rt]$ is the LO perturbative result in lattice theory 
with finite $a$ and $L$;
the third term is included for removing the remaining $\mathcal{O}(\alpha_s^2 a^2)$ effect.

We determine $\LMS$ in GeV units
by comparing the above corrected lattice data 
to the OPE prediction $V_S^{\rm RF}(r)+A_2 r^2$,
where each dataset is converted to GeV units using the estimated
$a_i^{-1} [\rm GeV]$.
In this global fit, there are 16 parameters in total:
$\LMS$, six $A_0$'s, $A_2$, six 
$\kappa$'s
and two $f$'s. 
Since we have more effective data than the first analysis,
we shift the fitting range to shorter distances.
It serves to reduce the higher order uncertainty, which is the dominant error in our analysis.
We use the lattice data at $\LMS^{\rm PDG} r<0.6$.
We obtain $\LMS=0.334 \pm 0.010 ({\rm stat}) ~{\rm GeV}$, giving $\alpha_s(M_Z)=0.1179 \pm 0.0007 ({\rm stat})$.\footnote{
The fit gives $\kappa_{d,i}$'s 
consistent with naively expected values $C_F \alpha_s(a_i^{-1})$;
$f_d$'s are consistent with zero.
}
For $A_2$, we have $A_2=-0.0091 \pm 0.0054$({\rm stat})~GeV$^3$.

We consider the following systematic errors.
Since our final result is obtained from Analysis (II), 
we consider systematics errors more in detail than in Analysis (I).
(II-a) Finite $a$ effect: We drop the data at 
$r <2 a$, while the data at $r\geqslant a$ are used 
in the main analysis.\footnote{
When dropping the data at $r <2 a$,  
we also drop the parameters $\kappa$
(effective for discretization effects by the
data at $r \sim a$).
It is because the roles of $\kappa$ and $f$ become degenerate
at larger $r$, which destabilizes the fit.
}
(II-b) Mass corrections: 
The input $(u,d,s)$  
masses in each lattice simulation differ from the physical point.
Since the nonperturbative correction due to these mass differences
is unknown,
we treat it as a systematic error.
The lattice data at the physical point are estimated using perturbation theory as
$V_{\rm latt}(r; m^{latt} ) \to V_{\rm latt}(r; m^{latt})+[V_{\rm pt}(r; \overline{m} )-V_{\rm pt}(r; m^{latt}) ]$,
where $V_{\rm pt}$ is a finite mass effect evaluated 
in perturbative QCD at 
${\rm N^2LO}$ \cite{Hoang:2000fm}
with the ${\rm \overline{MS}}$ masses $\overline{m}$. 
We also substitute a constituent quark mass of $300~{\rm MeV}$ for $\overline{m}$
to estimate the correction.
Furthermore, since in the main analysis we use $V_S^{\rm RF}$ in the massless approximation,
finite mass corrections are added.
(II-c) Higher-order corrections to $V_S^{\rm RF}$: An analysis parallel to the first one is performed.
(II-d) Matching range: The upper limit of $r$ is varied as $\LMS^{\rm PDG} r<0.8$ or
$0.5$. 
(II-e) US contributions: An analysis parallel to the first one is performed.
(II-f) Scheme dependence: The $\mf$-independent part of $V_S$ varies by
a choice of scheme.
A different scheme practically causes an $\mathcal{O}(r^3)$ difference 
in the OPE prediction (\ref{OPE}).
We add an $r^3$-term in the fit to remove the scheme dependence 
and see how $\alpha_s$ varies.
(II-g) Lattice spacing: The lattice spacing is shifted by its uncertainty.
We also take into account the error of the Wilson-flow scale. (See Ref.~\cite{prep} for details.)
We summarize the systematic errors in table~\ref{tab:systematics}.

\begin{figure}[h]
\vspace*{-2mm}
\begin{center}
\includegraphics[width=7cm]{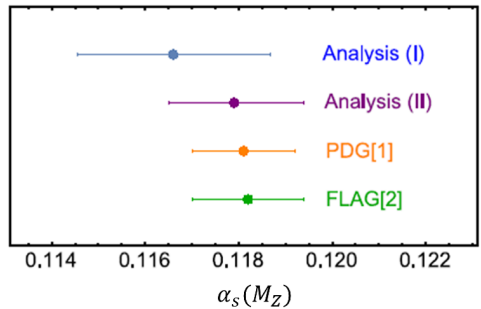}
\end{center}
\vspace*{-5mm}
\caption{
Comparison of 
$\alpha_s(M_Z)$ determinations.
The FLAG average is based on Refs.~\cite{Bazavov:2014soa,Chakraborty:2014aca}.
}
\label{Fig:alphas}
\vspace*{-5mm}
\end{figure}

As a result of the second analysis, we obtain
\be
\alpha_s(M_Z)=0.1179\pm 0.0007({\rm stat})^{+0.0014}_{-0.0012}({\rm sys}) \, .
\ee
We present the results of Analysis (I) and (II) in Fig.~\ref{Fig:alphas}, where one can see that
they are mutually consistent. 
Analysis (II) is superior to Analysis (I) in the sense that
it is a first-principle analysis and that our dominant error, higher order uncertainty,
is reduced thanks to the use of shorter distance range.
Hence, we adopt the result of Analysis (II) as our final result.

In this Letter
we determined $\alpha_s$ from the QCD potential by comparing 
the lattice result and OPE prediction after subtracting
renormalons from the leading Wilson coefficient.
We confirmed an agreement at $\LMS r \lesssim 0.8$, with good quality data
and consistent with expectation of OPE free of renormalons
(see Fig.~\ref{Fig:consistency}).
Consequently we obtained $\alpha_s(M_Z)=0.1179^{+0.0015}_{-0.0014}$
[from Analysis (II)].

The dominant error in this result stems from 
the uncertainty of the perturbative prediction.
Utilizing finer lattices will straightforwardly reduce
the error,
since the perturbative uncertainty decreases at smaller $r$. 

The authors are grateful to the JLQCD collaboration for providing
the lattice data.
This work is supported in part by Grant-in-Aid for
scientific research (Nos.\ 17K05404 and 26400255) from
MEXT, Japan.

\begingroup\raggedright\endgroup

\end{document}